\documentclass[12pt]{article}
\usepackage{latexsym,amssymb}
\newtheorem{definizione}{Definition}[section]
\newcommand{\bdefi}{\begin{definizione}}
\newcommand{\edefi}{\end{definizione}}
\makeatletter
\@addtoreset{equation}{section}
\makeatother

\newcommand{\ft}[2]{{\textstyle\frac{#1}{#2}}}
\newsavebox{\uuunit}
\sbox{\uuunit}
    {\setlength{\unitlength}{0.825em}
     \begin{picture}(0.6,0.7)
        \thinlines
        \put(0,0){\line(1,0){0.5}}
        \put(0.15,0){\line(0,1){0.7}}
        \put(0.35,0){\line(0,1){0.8}}
       \multiput(0.3,0.8)(-0.04,-0.02){10}{\rule{0.5pt}{0.5pt}}
     \def\bar{\overline}\end {picture}}

\begin{document}
\begin{titlepage}
\begin{flushright}
KUL-TF-2000/21\\
ULB-TH-00/18\\
hep-th/0009048 \\
\end{flushright}
\vskip 2cm
\begin{center}
{\Large \bf
Non-semisimple gaugings \\ of $D=5~~\mathcal{N}=8$ Supergravity
and FDA.s$^\dagger$\\
}
\vfill

{\large  Laura Andrianopoli$^{1,2}$, Francesco Cordaro$^1$,}\\
{\large Pietro Fr\'e$^3$ and Leonardo Gualtieri$^4$} \\
\vfill
{ \sl
$^1$  Instituut voor Theoretische Fysica - Katholieke Universiteit Leuven
\\Celestijnenlaan 200D B--3001 Leuven, Belgium\\
$^2$ Dipartimento di Fisica, Politecnico di Torino, corso Duca
degli Abruzzi 24, \\
I-10129 Torino, Italy\\
 $^3$ Dipartimento di Fisica Teorica, Universit\'a di Torino,
via P. Giuria 1,
I-10125 Torino, \\
 Istituto Nazionale di Fisica Nucleare (INFN) - Sezione di Torino,
Italy \\
$^4$ Physique Th\'eorique et Math\'ematique, Universit\'e Libre de
Bruxelles, C.P. 231, B--1050, Bruxelles, Belgium }
\end{center}
\vfill
\begin{abstract}
{We reformulate maximal $D=5$ supergravity in the consistent approach uniquely
based on Free Differential Algebras and the solution of their
Bianchi identities (= rheonomic method).
In this approach the lagrangian is unnecessary since the field equations
follow from closure of the supersymmetry algebra.
This enables us to explicitly construct the non--compact gaugings
corresponding to the non--semisimple algebras $\mathrm{CSO(p,q,r)}$,
irrespectively from the existence of a lagrangian.
The use of Free Differential Algebras
is essential to clarify, within a cohomological set up,
the dualization mechanism between one--forms and two--forms. Our theories
contain $12-r$ self--dual two--forms and $15+r$ gauge vectors, $r$ of which
are abelian and neutral.
These theories, whose existence is proved and their supersymmetry algebra
constructed hereby, have potentially interesting properties in relation with
domain wall solutions and the trapping of gravity.
}
\end{abstract}
\vspace{2mm} \vfill \hrule width 3.cm
{\footnotesize
 $^ \dagger $ \hskip 0.1cm Supported by   EEC  under TMR contract
 ERBFMRX-CT96-0045 and RTN contract RTN1-1999-00116}
\vspace{2mm} \vfill \hrule width 3.cm
\end{titlepage}
\section{Introduction}
Gauged supergravity with a maximal compact group, $\mathcal{G}=\mathrm{SO(6)}$ in $D=5$
\cite{gunwar,PPV}, $\mathcal{G}=\mathrm{SO(8)}$ in
$D=4$ \cite{dewitnicol} and $\mathcal{G}=\mathrm{USp(4)}$ in $D=7$ \cite{Pernici:1984xx}
has attracted much renewed attention in the last two
years because of the $\mathrm{AdS_{p+2}/CFT_{p+1}}$ correspondence (for a general review
see \cite{Aharony:2000ti} and references therein; for the case $D=p+2=4$ see also
\cite{Gualtieri:1999tu} and references therein). Indeed the
maximally supersymmetric vacuum of these gauged supergravities is the $AdS_D$
space and the compact gauge group $\mathcal{G}_{gauge}$ is the
$\cal{R}$-symmetry of the corresponding maximally extended supersymmetry algebra.
\par
However the compact gaugings are not the only ones for extended
supergravities. There exist also versions of these theories where the gauge
group  $\mathcal{G}_{gauge}$ is non--compact. Unitarity is preserved because
in all possible extrema of the corresponding  scalar potential the
non--compact gauge symmetry is broken to some residual compact subgroup.
Furthermore, there are models in which the gauge group is non--semisimple.
For ${\cal N} =8$ in $D=4$, they were particularly studied by Hull \cite{hull,hull2}
and an exhaustive classification of these  gaugings
was more recently obtained  by some of us \cite{noi4D}.
\par
The non--semisimple gauged supergravities are  relevant for a close relative of the
$\mathrm{AdS/CFT}$ correspondence namely the
\begin{equation}
  \mbox{Domain Wall}/QFT \,\,\, \mbox{correspondence}
\label{domwalqft}
\end{equation}
between gauged supergravities and quantum field theories realized on domain wall
solutions of either string theory or M--theory. This generalization of the Maldacena
conjecture was introduced by Boonstra, Skenderis and Townsend \cite{Boonstra:1999mp}
and has been further developed in recent times \cite{Cvetic:2000pn,Cvetic:2000ah}.
Indeed after the challenging proposal by Randall and Sundrum \cite{Randalsud}
that {\it compactification of extra dimensions} can be traded for the {\it trapping of
gravity} on $4$--dimensional branes, much interest has gone into finding
supergravity theories that can accomodate the Randall Sundrum scenario
\cite{Ceresole:2000jd}, \cite{newduffo} . These
have been related to domain--walls in \cite{Cvetic:2000pn},
and hence to non--semisimple gauged supergravities \cite{Boonstra:1999mp}.
\par
For all these reasons it is interesting to study the
non--semisimple gaugings of $D=5$ supergravity, both in the case
of lower and maximal supersymmetry. For maximal $ \mathcal{N}=8$
supergravity in five--dimensions the analogue of the $D=4$
exhaustive classification derived in \cite{noi4D} has not been
obtained so far.  G\"unaydin, Romans and Warner have constructed
the $\mathrm{SO(6-q,q)}$ gaugings \cite{gunwar} that are the
analogues of the $\mathrm{SO(8-q,q)}$ gaugings in four--dimensions
but so far no gauging based on the so called $\mathrm{CSO(p,q,r)}$
contracted algebras (with $p+q+r=6$) has been produced. These
gaugings exist in $4$--dimensions (with $p+q+r=8$) and it would be
natural to assume that they also exist in $5$--dimensions. The
difficulty met by the authors who have so far investigated this
problem resides in the novel five--dimensional feature of
one--form/two--form duality.  As long as all vector fields are
abelian we can consider them as one--form or two--form gauge
potentials at our own will. Yet when we introduce a certain degree
of non--abelian gauge symmetry matters become more complicated,
since only $1$--forms can gauge non--abelian groups while
$2$--forms cannot. On the other hand $1$--forms that transform in
a non--trivial representation of a non abelian gauge group which
is not the adjoint representation are equally inconsistent. They
have to be replaced by $2$--forms and some other mechanism,
different from gauge symmetry has to be found to half their
degrees of freedom. This is self--duality between the $2$--form
and its field strength. Hence gauged supergravity can only exist
with an appropriate mixture of $1$--forms and self--dual
$2$--forms. While this mixture was mastered in the case of compact
and non--compact but semisimple gaugings, the case of
$\mathrm{CSO(p,q,r)}$ algebras that are not semisimple seemed to
be unreachable in the existing literature.
\par
In the present paper we show that the $\mathrm{CSO(p,q,r)}$ gaugings do exist and
are fairly simple. The catch is the use of the geometric approach (based on Free
Differential Algebras \footnote{for  comprehensive
reviews of these topics see vol.2 of
\cite{castdauriafre}}) where the mechanism of one--form/two--form dualization
receives a natural algebraic formulation and explanation.
\par
The final result is that in the case of the $\mathrm{CSO(p,q,r)}$ gaugings there
are $15+r$ gauge vectors and $12-r$ self--dual two--forms. $15$ of the
vectors gauge the contracted algebra while $r$ of them have an abelian gauge symmetry with
respect to which no field in the theory is charged. At the same time these vectors are
neutral with respect to the transformations of the gauge algebra. Furthermore
how many fields are true vectors and how many are replaced by self--dual
two--forms is decided by a
cohomological argument clearly formulated in the Free Differential Algebra set up.
\par
The price one might be forced to pay in the case of $r>0$ extra
neutral vector fields is that, although field equations can be
normally derived from closure of the supersymmetry algebra, yet a
lagrangian of conventional type might not exist, just as it
happens for type $IIB$ supergravity in $D=10$ (after all, this is
not terribly surprising since ${\cal N}=8$ supergravities in five
dimensions should eventually be interpreted in terms of brane
mechanisms and compactifications from type IIB superstring).
This whould make impossible the construction of the theory
by means of lagrangian--based techniques.
However in our construction, based on the closure of the
supersymmetry algebra, the existence of a Lagrangian is not
fundamental, the existence of the theory following from the
consistent closure of Bianchi-Identities.

The scalar potential of these supergravities can be systematically
derived, together with the complete field equations, from the closure of the
supersymmetry algebra we have determined in the present paper.
This is completely algorithmic and straightforward, but it involves lengthy
calculations
that are postponed to a forthcoming publication \cite{next},
where a full--fledged investigation of the properties of the potential and of its
implications for the correspondence (\ref{domwalqft}) will be given.
\section{$D=5~~\mathcal{N}=8$ supergravity}
In this section we recall the main features of
$D=5~~\mathcal{N}=8$ supergravity theory \cite{gunwar},
\cite{PPV}, fitting its formulation into the framework of the
rheonomic constructions \cite{castdauriafre} and of the general
discussion of duality symmetries \cite{mylecture} and central
charges \cite{Andrianopoli:1997wf}. While adopting where possible
the conventions of \cite{gunwar}, recasting  $D=5~~\mathcal{N}=8$
supergravity into the general framework of
\cite{Andrianopoli:1997wf} is also a matter of notations since the
names given to the various types of indices must reflect their
interpretation within the framework. Specifically the notations
are as follows. By $A,B=1,\dots,8$ we denote the indices labeling
the supercharges and acted on by the {\it isotropy} subgroup $H$
of the scalar coset $G/H$. In our case they are in the fundamental
${\bf 8}$ of $\mathrm{USp(8)}$. The indices
$\Lambda,\Sigma=1,\dots,27$  label instead the vector fields and
correspond to the linear representation of the scalar {\it
isometry} group $G$ to which the vectors are assigned. In our case
they run in the ${\bf 27}$ of $\mathrm{E_{6(6)}}$. Next we need a
notation for the electric subgroup $\mathrm{SL}(6,\mathbb{R})
\times \mathrm{SL}(2,\mathbb{R}) \subset \mathrm{E_{6(6)}}$ within
which the generators of the gauge group can be chosen. It is as
follows. $I,J=1,\dots,6$ are indices in the fundamental  ${\bf 6}$
both of $\mathrm{SL}(6,\mathbb{R})\subset \mathrm{E_{6(6)}}$ and
of $\mathrm{SO(6)}$ (or of its non--compact/non--semisimple
versions); the indices $\alpha,\beta=1,2$ run in the fundamental
$\bf 2$ of $\mathrm{SL}(2,\mathbb{R})\subset \mathrm{E_{6(6)}}$.
Finally $\mu,\nu=0,\dots,4$ are the usual curved spacetime
indices, while  we call $a,b=0,\dots,4$ the flat indices of the
f\"unfbein. The conventions for the gamma matrices, the spinors
and the symplectic metric as those used in \cite{gunwar} and
\cite{cremmer}.
\subsection{The ungauged theory}
The supersymmetry algebra for the ungauged theory is the
superPoincar\'e superalgebra, whose external automorphism
symmetry (the $\mathcal{R}$--symmetry) is $\mathrm{USp(8)}$. The
theory is invariant under local $\mathrm{ISO(4,1)}\times \mathrm{USp(8)}$ and
global $\mathrm{E_{6(6)}}$ transformations,
and under local supersymmetry transformations, generated by $32$ real
supersymmetry charges, organized in the eight pseudo--Majorana spinors
\begin{equation}
\mathrm{Q}^A=\Omega^{AB} \, \mathcal{C} \, \left( \overline{\mathrm{Q}}_A\right) ^T \,
 ~~~A=1,\dots,8\,.
\end{equation}
Here $\Omega^{AB}=-\Omega^{BA}$ is the $\mathrm{USp(8)}$ invariant
metric and $ \mathcal{C}$ is the $5$--dimensional charge conjugation matrix.
The theory contains: the graviton field, namely the f\"unfbein 1--form $V^a$,
eight gravitinos $\psi^A \equiv \psi^A_\mu \, dx^\mu$ in the
$\bf 8$ representation of $\mathrm{USp(8)}$, $27$ vector fields
 $A^{\Lambda} \equiv A^{\Lambda} _\mu \, dx^\mu$ in the $\bf 27$ of
 $\mathrm{E_{6(6)}}$,
$48$ dilatinos $\chi^{ABC}$ in the $\bf 48$ of $\mathrm{USp(8)}$, and $42$ scalars
$\phi$ that
parametrize the coset manifold $E_{\left(6\right)6}/\mathrm{USp(8)}$, and
appear in the theory through the coset representative $\mathbb{L}_{\Lambda}^{~AB}(\phi)$, in the
$\bf (27,\overline{{27}})$ of $\mathrm{USp(8)}\times \mathrm{E_{6(6)}}$.
The local $\mathrm{USp(8)}$ symmetry is gauged by the composite connection
built out of the scalar fields. The connection (in the ${\bf 36}$ of $\mathrm{USp(8)}$)
and the vielbein (in the ${\bf 42}$ of $\mathrm{USp(8)}$) of the
scalar manifold are defined through the following relation:
\begin{equation}
\mathbb{L}^{-1~\Lambda}_{AB}d\mathbb{L}_{\Lambda}^{~CD}=
\mathcal{Q}_{[A}^{~~~[C}\delta^{D]}_{B]}+\mathcal{P}_{AB}^{~~~CD}\,.
\label{defungaugedconnection}
\end{equation}
The isometry of the scalar manifold, $E_{\left(6\right)6}$, is a global symmetry of
the theory.
\subsection{The gauging}
In maximal supergravities, where no matter multiplets can be added,
{\it gauging} corresponds to the addition of suitable
interaction terms that turn  a subgroup $\mathcal{G}$ of the global
$E_{\left(6\right)6}$ duality group into a local symmetry.
This is done by means of vectors chosen among the $27$ $A^{\Lambda}$.
The $E_{\left(6\right)6}$ symmetry is broken to the normalizer of
$\mathcal{G}$ in
$E_{\left(6\right)6}$, and after this operation
 the new theory has a local symmetry $\mathrm{USp(8)}\times\mathcal{G}$
and a global symmetry $N(\mathcal{G},E_{\left(6\right)6})$.
The choice of $\mathcal{G}$ is strictly constrained by the request that the vectors which
gauge this symmetry should transform in the coadjoint representation of $\mathcal{G}$,
so that the following branching must be true:
\begin{equation}
\label{gaugprop}
{\bf 27}\stackrel{\mathcal{G}\subset E_{\left(6\right)6}}{\longrightarrow}
Coadj\left(\mathcal{G}\right)\oplus\hbox{~rep. of~}\mathcal{G}\,.
\end{equation}
It turns out that this request is satisfied if and only if
$\mathcal{G}$ is a fifteen--dimensional
subgroup of $\mathrm{SL}(6,\mathbb{R})\subset E_{\left(6\right)6}$ whose adjoint
is identified with the $\bf 15$ representation of
$\mathrm{SL}(6,\mathbb{R})$. Indeed the $\bf 27$ of $E_{\left(6\right)6}$ decomposes under
\begin{equation}
\mathrm{SL}(6,\mathbb{R})\times \mathrm{SL}(2,\mathbb{R})\subset E_{\left(6\right)6}
\end{equation}
as
\begin{equation}
\bf 27 \longrightarrow ({\bar{15}},1)\oplus (6,2)\,
\end{equation}
(for example, $\mathbb{L}_{\Lambda}^{~AB}\longrightarrow
 (\mathbb{L}^{IJAB},\mathbb{L}_{I\alpha}^{~AB})$)
so that the property (\ref{gaugprop}) is satisfied. The subgroups
of $\mathrm{SL}(6,\mathbb{R})$ whose adjoint is the $\bf 15$ of
$\mathrm{SL}(6,\mathbb{R})$ are the $\mathrm{SO(p,q)}$ groups with
$p+q=6$ and their contractions $\mathrm{CSO}\mathrm{(p,q,r)}$,
which will be discussed in section $4$ (see
\cite{hull,hull2,noi4D} for definitions). The possible gaugings
are then restricted to these groups. The normalizer in
$E_{\left(6\right)6}$ of all these groups is the same as the
normalizer of $\mathrm{SL}(6,\mathbb{R}) $, namely
 $\mathrm{SL}(2,\mathbb{R})$. Therefore this latter is the
residual global symmetry for all possible gaugings. The 27 vectors
$A^{\Lambda}$ are then decomposed into the vectors $A_{IJ}$ in the
$\bf (\bar{15},1)$, that gauge $\mathcal{G}$, and the vectors  in
the $\bf (6,2)$, which do not gauge anything and are then forced
(as we will see later) to be dualized into two--forms
$B^{I\alpha}$. The fifteen generators $G^{IJ}$ of $\mathcal{G}$
can be expressed as linear combinations of the $35$ generators
$G_\ell$ ($\ell\!\!=\!\!1,\dots 35$) of
$\mathrm{SL}(6,\mathbb{R})$: $G^{IJ}\!=\!G_\ell \,e^{\ell IJ}$
where $e^{\ell IJ}$ is the {\it embedding matrix} \cite{noi4D}
which describes the embedding of $\mathcal{G}$ into
$\mathrm{SL}(6,\mathbb{R})$. For all the admissible cases  in the
fundamental $\bf 6$--dimensional representation the generators of
the gauge group $\mathcal{G}$ take the form \cite{gunwar}
\begin{equation}
(G^{IJ})^K_{~L}=\delta^{[I}_L\eta^{J]K}
\label{genfund}
\end{equation}
where $\eta^{JK}$ is a diagonal matrix with $p$ eigenvalues equal to $1$,
 $q$ eigenvalues equal to $(-1)$   and, only in the case of contracted groups,
  $r$  null eigenvalues. This signature completely characterizes
the gauge groups and correspondingly the
gauged theory.
From (\ref{genfund}) one can build the generators of
$\mathcal{G}\subset E_{\left(6\right)6}$ in the
$\bf 27$ representation of $E_{\left(6\right)6}$, namely
some suitable matrices $\,(G^{IJ})_{\Lambda}^{~\Sigma}$.
According to the general framework of
\cite{mylecture,castdauriafre,noi4D}, in presence of gauging,
the composite $H$--connection of $\mathrm{USp(8)}$ and scalar vielbein,
defined in (\ref{defungaugedconnection}) are replaced by their gauged analogues:
\begin{equation}
\mathbb{L}^{-1~\Lambda}_{AB}d\mathbb{L}_{\Lambda}^{~CD}+
g(\mathbb{L}^{-1})_{AB}^{~\Lambda}(G^{IJ})_{\Lambda}^{~\Sigma}\mathbb{L}_{\Sigma}^{~CD}
A_{IJ}=\hat{\mathcal{Q}}_{[A}^{~~~[C}\delta^{D]}_{B]}+\hat{\mathcal{P}}_{AB}^{~~~CD}\,,
\label{defgaugedconnection}
\end{equation}
where $g$ is the gauge coupling constant.
The covariant $\mathrm{USp(8)}$ derivative of a field $V_A$
is defined as
\begin{equation}
\nabla V_A=\mathcal{D} V_A+\hat{\mathcal{Q}}_A^{~B}\wedge V_B
\label{covQder}
\end{equation}
where $\mathcal{D}$ is the Lorentz--covariant exterior derivative.
The covariant derivative with respect to $\mathcal{G}$
of a field $V^I$ in the $\bf 6$ of $ \mathrm{SL(6, \mathbb{R})}$ is instead defined as
follows:
\begin{equation}
DV^I\equiv \nabla V^I+g(G^{KL})^I_{~J}A_{KL}\wedge V^J\,.
\label{covGder}
\end{equation}
The field content of the gauged supergravity theory is the following
\begin{equation}
\begin{array}{|c|c|c|c|c|}
\hline
\#& \hbox{Field} & \left(SU(2)\times SU(2)\right) \hbox{--spin~rep.} & \mathrm{USp(8)}
\hbox{~rep.} & \mathcal{G}
\hbox{~rep.} \\
\hline
1& V^a & (1,1) & {\bf 1} & {\bf 1} \\
\hline
8& \psi^A & (1,1/2)\oplus (1/2,1) & {\bf 8} & {\bf 1} \\
\hline
15& A_{IJ} & (1/2,1/2) & {\bf 1} & {\bf 15} \\
\hline
12& B^{I\alpha} & (1,0)\oplus (0,1) & {\bf 1} & {\bf 6\oplus \overline{{6}}} \\
\hline
48& \chi^{ABC} & (1/2,0)\oplus (0,1/2) & {\bf 48} & {\bf 1} \\
\hline
42& \mathbb{L}_{\Lambda}^{~AB}\left(\phi\right) & (0,0) & {\bf 27} & {\bf\overline{{27}}} \\
\hline
\end{array}
\label{gaugedfieldcontent}
\end{equation}
\section{Gauged supergravities from
Free Differential Algebras  and Rheonomy} Gauged maximal
supergravities in $D=5$ were originally constructed within the
framework of  No\"ether coupling and component formalism
\cite{gunwar},\cite{PPV}. As we pointed out in the introduction
the gaugings corresponding to the contracted groups
$\mathrm{CSO(p,q,r)}$ were left open in that approach. We are able
to construct explicitly all these theories by reverting to our
preferred approach based on Free Differential Algebras (FDA.s) and
the principle of rheonomy \cite{castdauriafre}. Indeed, within
this approach the non-semisimple theories are shown to exist and
explicitely constructed irrespectively on the existence or not of
a lagrangian formulation. Moreover, all the subtle points
concerning the role of two--form dualization are naturally
resolved in the Free Differential Algebra rheonomic approach. As
far as five dimensions are concerned this was already noted  in
\cite{Ceresole:2000jd} where the hypermultiplets were coupled to
$\mathcal{N}=2$ supergravity. Similarly the essential role of
FDA.s in gauging theories with $p$--form gauge fields was made
evident in \cite{F4} where the unique six--dimensional
$F(4)$--supergravity was finally constructed.
\subsection{The rheonomy principle}
For completeness, let us briefly recall the main steps in the
{\sl ''rheonomy approach''} to supergravity.
The starting point is to consider as fundamental fields the set of $1$--forms
$\mu^A\equiv\{\omega^{ab},\psi^{\alpha}, V^a\}$, that constitute
a cotangent frame dual to the Poincar\'e super--Lie algebra generators
$\{J_{ab},Q_{\alpha},P_a\}$.
The ordinary space--time parametrized by $\{x^{\mu}\}$ coordinates
can be extended to a superspace
parametrized also by the fermionic  $\{\theta^{\alpha}\}$ spinor coordinates.
We can give to the space--time fields
$\mu^A(x)$ a $\theta$-dependence through an appropriate extension mapping:
\begin{equation}
\label{rheomap}
\left\{
\begin{array}{lcl}
\omega^{ab}(x) &\rightarrow& \omega^{ab}(x,\theta) \\
\psi^{\alpha}(x) &\rightarrow& \psi^{\alpha}(x,\theta) \\
V^{a}(x) &\rightarrow& V^{ab}(x,\theta) \,.
\end{array}
\right.
\end{equation}
In such a way the bosonic space-time fields $\mu^A(x)$ are the
boundary values at $\theta^{\alpha}=0$ of these superspace fields
\begin{equation}
\mu^A(x)\equiv \mu^A(x,\theta)| _{\theta =d\theta =0}\,.
\end{equation}
The same extension holds also  for the set of curvature $2$--forms defined through the
structural equations:
\begin{equation}
R^A(x,\theta)\equiv d\mu^A(x,\theta)+{\ft 1 2} C^A_{~BC}\mu^B(x,\theta)\wedge\mu^C(x,\theta)
\equiv R^A_{LM}dZ^L\wedge dZ^M
\label{maucartcurv}
\end{equation}
that generalize Maurer--Cartan equations obtained by setting $R^A=0$.
In eq. (\ref{maucartcurv})
$R^A$ denotes the multiplet $\{R^{ab},\rho,R^a\}$ of super--Poincar\'e curvatures
and $dZ^L\equiv\{dx^{\mu},d\theta^{\alpha}\}$ is the set of coordinates which
span the cotangent space to superspace.
\par
In order to be completely determined as functions of
$x^{\mu},\theta^{\alpha}$, the fields $\mu^A$  must be equipped
with a complete set of Cauchy boundary conditions, namely we have to specify both
the space--time configurations $\mu^A_{\mu}(x,0)$ on the boundary $\theta =0$
and  the first--order derivatives along the theta
directions $\partial_{\theta^{\alpha}}\mu^A_{\mu}(x,\theta)|_{\theta =0}$
on the same boundary.
These derivatives can be expressed in terms of  $\theta$ projections
$R^A_{\alpha,L}$ of the $R^A$ curvatures.
The {\it extension map} (\ref{rheomap}) can thus be determined by
specifying the following two sets of boundary
values:
\begin{equation}
\mu^A(x,0) \quad ; \quad R^A_{\alpha,L}(x,0)\,.
\end{equation}
The former are the space--time configurations for the fields
$\{\omega^{ab}(x)$, $\psi^{\alpha}(x)$, $V^a(x)\}$. In order to determine the latter,
namely the so called {\sl outer} components of
the curvatures,
we make use of the {\sl rheonomy principle}, which states that the
outer
components of the curvatures are linear combinations of the {\sl inner} ones, i.e. of the
space--time configurations $\{R^{ab}(x),\rho(x),R^a(x)\}$:
\begin{equation}
\label{rheoconstraints}
R^A_{\alpha,L} = \mathcal{C}^{A|\mu\nu}_{\alpha L|B}R^B_{\mu\nu}
\end{equation}
where the $\mathcal{C}$'s are suitable constant tensors.
This expansion is called the {\sl  rheonomic parametrization}
of the curvatures.
The values of the constants $\mathcal{C}$ can be determined by imposing the closure of
Bianchi identities
\begin{equation}
dR^A+C^A_{BC}\mu^B\wedge R^C=0\,.
\end{equation}
Since by definition we have:
\begin{equation}
R^A_{\mu\nu}=\partial_{[\mu}\mu^A_{\nu]}+{\ft 1 2} C^A_{~BC}\mu^B_{[\mu}\wedge\mu^C_{\nu]}
\end{equation}
it turns out that the knowledge of the pure space-time configurations
$\{\!\mu_{\nu}^A(x,0),\!\partial_{\mu}\mu^A_{\nu}(x,0)\!\}$
completely determines the superspace extensions defined in (\ref{rheomap}).
It is worth noting that in this context  Bianchi
identities are not identically satisfied. This
is not surprising since supersymmetry is an on--shell symmetry,
and therefore it closes only modulo the equations of motion.
Bianchi identities are actually
equations of the theory, determining its dynamics.  Not only they give the
rheonomic parametrizations, but they also fix the geometry of the scalar manifold
and give the equations of motion satisfied by the spacetime fields.
In this framework, a supersymmetry transformation of the fields $\mu^A$ is
given by a diffeomorphism along a fermionic direction in superspace
$\varepsilon=\varepsilon^{\alpha}\partial_{\alpha}$ and is expressed
by means of a Lie derivative along $\varepsilon$:
\begin{equation}
\delta\mu^A(x,\theta)=l_{\varepsilon}\mu^A(x,\theta) = (\nabla\varepsilon)^A
+ 2 \varepsilon^{\alpha}\mathcal{C}^{A|\mu\nu}_{\alpha L|B}R^B_{\mu\nu}dZ^L
\label{lie}
\end{equation}
from which one can retrieve the  supersymmetry transformation rules of the
fields $\delta\psi_{\mu},\dots$ as given
in the usual component formalism. Note that the Lie derivatives $l_{\varepsilon}$
close a super--Lie algebra, namely:
\begin{equation}
\left[l_{\varepsilon_1},l_{\varepsilon_2}\right] = l_{[\varepsilon_1,\varepsilon_2]}
\end{equation}
if the integrability condition $d^2=0$ is used. Of course this requirement is
equivalent to enforcing the
closure of  Bianchi identities for the curvatures $R^A$.
\par
Summarizing, to construct a supergravity theory we use the rheonomic conditions
(\ref{rheoconstraints}) for the curvatures and then we solve Bianchi
Identities in two steps.
In the {\em first step} we analyse the sectors that determine the unknown
coefficients $\mathcal{C}$'s.
Once a set of $\mathcal{C}$'s satisfying Bianchi Identities has
been found, the corresponding supergravity theory has been {\em proven to
  exist} and its on--shell closed supersymmetry algebra has been constructed.
Indeed, we have an explicit and consistent form for all the susy transformation rules.
\par
Moreover, the classical dynamics of the theory is completely
determined, since the classical equations of motion uniquely follow from closure of the susy
algebra.
To work them out explicitly corresponds to the {\em second step}.
It suffices to complete the analysis of B-I in the remaining sectors
by means of calculations that are completely straightforward and guaranteed, although somewhat lengthy.
From this viewpoint, the explicit construction of the Lagrangian $\cal L$ is not really needed.
Simply, when $\cal L$ exists, the determination of the field equations is more
easily obtained by $\delta\cal L$ variations than through the analysis of
the remaining sectors of Bianchi Identities
\footnote{Conceptually, the lagrangian is essential only for quantization, which
however, in a modern perspective, is not the issue for theories, like
supergravity, that are regarded as effective low energy theories of more
fundamental microscopic quantum theories, like string theory.}.
When the Lagrangian exists, it can be obtained by means of a straightforward
procedure starting from the rheonomic parametrizations \cite{castdauriafre}.
\par
What we sketched above describes minimal supergravity, containing only the
graviton, the gravitino and the spin connection, but it is easily generalized to
all supergravity theories \cite{castdauriafre}, where also other fields are present, such as dilatinos,
scalars, vectors and higher order forms. In these cases, the whole
construction can be repeated with the $\mu^A$ defined to include all the
$1$--forms of the theory. Scalars and spin--one half fields are introduced by
including their covariant derivatives as additional curvatures of the
theory. When higher order forms are present, the super-Lie algebra has to be
enlarged to a Free Differential Algebra expressing the occurrence of a
higher order cohomology.
\subsection{Rheonomic parametrizations for gauged $\mathcal{N}=8$  supergravity
in five dimensions}
In the theory we are considering, the relevant curvatures are defined below:
\begin{equation}
  \begin{array}{rl}
    \left . \mbox{\vbox{\hbox{Poincar\'e}
  \hbox{2-form}\hbox{curvatures}}} \quad \right \{ & \begin{array}{rcl}
 R^a &\equiv& \mathcal{D} V^a +{{\rm i}\over 2} \bar{\psi}^A\wedge\gamma^a\psi_A \\
R^{ab} &\equiv& d\omega ^{ab} - \omega ^a_{~c}\wedge\omega ^{cb}  \\
\rho_A &\equiv& \nabla\psi_A  \
\end{array}  \\
\null & \null \\
   \left . \mbox{\vbox{\hbox{vector}
  \hbox{2-form}\hbox{curvature}}}  \quad\right \{ & \begin{array}{rcl}
 F_{IJ} &\equiv& dA_{IJ}+\frac{1}{2} g f_{IJ}^{~~~KL,MN}A_{KL}\wedge A_{MN}\\
  \null & \null & -
{\rm i} \mathbb{L}^{-1}_{CDIJ}\bar{\psi}^C\wedge\psi^D \
\end{array}  \\
   \left . \mbox{\vbox{\hbox{dilatino}
  \hbox{1-form}\hbox{curvature}}}  \quad\right \{ & \begin{array}{rcl}
  X_{ABC} &\equiv& \nabla\chi_{ABC} \
\end{array}  \\
   \left . \mbox{\vbox{\hbox{composite}
  \hbox{1-form}\hbox{$\mathrm{USp(8)}$}\hbox{curvature}}}  \quad\right \{  &
  \begin{array}{rcl}
  \hat{R}_A^{~B} &\equiv& d\hat{\mathcal{Q}}_A^{~B} +
  \hat{\mathcal{Q}}_A^{~C}\wedge\hat{\mathcal{Q}}_C^{~B}\
\end{array}  \\
   \left . \mbox{\vbox{\hbox{3--form}
  \hbox{curvatures}\hbox{of the}\hbox{2--forms}}}  \quad \right \{  & \begin{array}{rcl}
  H^{I\alpha} &\equiv&
D\left[B^{I\alpha}+{\rm
i}\mathbb{L}^{-1~I\alpha}_{AB}\bar{\psi}^A\wedge\psi^B\right]\
\end{array} \
  \end{array}
\label{defcurvatures}
\end{equation}
where $\nabla$, $D$ denote the complete covariant differentials
according to eq.s (\ref{covQder}), (\ref{covGder}) and the
Lorentz--covariant
derivatives of the vielbein and the gravitino
$1$--forms are defined below:
\begin{eqnarray}
\mathcal{D} V^a &\equiv& dV^a-\omega ^a_{~b}\wedge V^b \nonumber \\
\mathcal{D}\psi_A &\equiv& d\psi_A - \frac{1}{4}\omega ^{ab}\gamma_{ab}\wedge\psi_A
\end{eqnarray}
$\omega ^{ab}$ being the spin connection. With $f_{IJ}^{KL,MN}$ we have denoted
the structure constants of the gauge group $\mathcal{G}$.
The curvatures (\ref{defcurvatures}) satisfy the following set of Bianchi identities:
\begin{equation}
\begin{array}{l}
\mathcal{D} R^a + R^a_{~b}\wedge V^b -{\rm i} \bar{\rho}^A\wedge\gamma^a\psi_A = 0 \nonumber\\
\mathcal{D} R^{ab} = 0 \nonumber\\
\nabla\rho _A + \frac{1}{4} R^{ab}\wedge\gamma_{ab}\psi_A - \hat{R}_{A}^{~~B}\wedge\psi_B =  0 \nonumber \\
\nabla X_{ABC} + \frac{1}{4} R^{ab}\gamma_{ab}\chi_{ABC} - 3
R_{[A}^{~~D}\wedge\chi_{BC]D} =  0 \nonumber \\
DF_{IJ} - {\rm i}\mathbb{L}^{-1}_{CDIJ}\left(\hat{P}^{CDEF}\wedge\bar{\psi}_E\wedge\psi_F
- 2 \bar{\rho}^C\wedge\psi^D\right) = 0 \nonumber\\
DH^{I\alpha} - g (G^{KL})^I_{~J}\left[F_{KL}+{\rm i}\mathbb{L}^{-1}_{ABKL}\bar{\psi}^A\wedge\psi^B\right]\wedge
\left[B^{J\alpha}+{\rm i}\mathbb{L}^{-1~J\alpha}_{CD}\bar{\psi}^C\wedge\psi^D\right] = 0\nonumber\\
\hat{R}^A_{~B}=-{1\over 3}\hat{P}_{ACDE}\wedge\hat{P}^{BCDE}+{1\over 3}gT^B_{~~AEF}(\phi)
(\mathbb{L}^{IJEF}F_{IJ}+{\rm i}{\bar \psi}^E\wedge\psi^F) \nonumber\\
\nabla\hat{P}_{ABCD}=gY^+_{ABCDEF}(\phi)(\mathbb{L}^{IJEF}F_{IJ}+{\rm i}{\bar \psi}^E\wedge\psi^F)
\nonumber\\
\end{array}
\label{bianchi}
\end{equation}
where the $\mathrm{USp(8)}$--tensors $T^B_{~~AEF}(\phi)$, $Y^+_{ABCDEF}(\phi)$ are defined as follows:
\begin{eqnarray}
Y^{AB}_{~~~CDEF}&\equiv&
\mathbb{L}^{-1~~\Lambda}_{~CD}(G^{IJ})_{\Lambda}^{~\Sigma}\mathbb{L}_{\Sigma}^{~AB}
\mathbb{L}^{-1}_{EFIJ}\\\nonumber\\
Y^{\pm}_{ABCDEF}&\equiv&\frac{1}{2}(Y_{ABCDEF}\pm Y_{CDABEF})\\\nonumber\\
T^A_{~~BCD}&\equiv&Y^{AF}_{~~~BFCD}\,.
\label{YTtensors}
\end{eqnarray}
The solution to the Bianchi identities
(that is, the rheonomic parametrization) is given,
modulo bilinears in the dilatinos, by  the following expressions in terms
of the inner components $R^{ab}_{~~cd}, \rho^A_{ab}, \dots$:
\begin{eqnarray}
R^a &=& 0 \\
R^{ab} &=& {1\over 2}
     R^{ab}_{~~cd} V^c\wedge V^d + \nonumber\\\nonumber\\
&& {2\over 3}{\rm i} \mathcal{H}_{AB}^{ab}\bar{\psi}^A\wedge\psi^B
   + {1\over 6}{\rm i} \mathcal{H}_{AB|cd}\bar{\psi}^A\wedge\gamma_e\psi^B\epsilon^{abcde} +
     g {2\over 45}{\rm i}T_{AB}\bar{\psi}^A\wedge\gamma^{ab}\psi^B  \nonumber\\\nonumber\\
&& - 2{\rm i} \bar{\rho}^{A~c[a}\gamma^{b]}\psi_A\wedge V_c +
     {\rm i} \bar{\rho}^{A~ab}\gamma^c\psi_A \wedge V_c + \mathcal{O} (\chi^2)\\
\rho _A  &=&
      \rho _{A|ab} V^a\wedge V^b - g {2\over 45}T_{AB}\gamma_a\psi^B\wedge V^a  \nonumber\\\nonumber\\
&& - {2\over 3} \mathcal{H}_{AB|ab}\gamma^a\psi^B\wedge V^b +
     {1\over 12} \mathcal{H}_{AB|ab}\gamma_{cd}\psi^B\wedge V_e \epsilon^{abcde}  \nonumber\\\nonumber\\
&& + {3{\rm i}\over 4\sqrt{2}} \chi_{ABC}\bar{\psi}^B\wedge \psi^C -
{{\rm i}\over 4\sqrt{2}} \gamma_a\chi_{ABC} \bar{\psi}^B\gamma^a\wedge\psi^C + \mathcal{O} (\chi^2) \\
\nabla\chi_{ABC} &=&
      (\nabla_a\chi_{ABC})V^a + {1\over\sqrt{2}} gA^D_{ABC}\psi_D +
\sqrt{2} \hat{P}_{ABCD|a}\gamma^a\psi^D \nonumber\\\nonumber\\
&& - {3\over 2\sqrt{2}}\mathcal{H}_{[AB|ab}\gamma^{ab}\psi_{C]}
-{1\over 2\sqrt{2}} \Omega_{[AB}\mathcal{H}_{C]D|ab}\gamma^{ab}\psi^D +
\mathcal{O} (\chi^2) \\
F_{IJ} &=& {1\over 2}F_{IJ|ab}V^a\wedge V^b + {{\rm i}\over\sqrt{2}}
\mathbb{L}^{-1}_{ABIJ}\bar{\chi}^{ABC}\gamma_a\psi_C\wedge V^a \\\nonumber\\
B^{I\alpha} &=& {1\over 2}B^{I\alpha}_{ab}V^a\wedge V^b +
{{\rm  i}\over\sqrt{2}}\mathbb{L}^{-1~I\alpha}_{AB}\bar{\chi}^{ABC}\gamma_a\psi_C\wedge V^a
\label{rheoB}\\
H^{I\alpha} &=& H^{I\alpha}_{abc}V^a\wedge V^b\wedge V^c +
g {{\rm i}\over 2}\mathbb{L}^{I\alpha}_{~~AB}\bar{\psi}^A\wedge\gamma_a\psi^B\wedge V^a
\nonumber\\\nonumber\\
&&-g {{\rm i}\over 4\sqrt{2}}
\mathbb{L}^{I\alpha}_{~~AB}\bar{\chi}^{ABC}\wedge\gamma_{ab}\psi_C\wedge V^a\wedge V^b
\label{rheoH}\\
\hat{P}^{ABCD} &=& \hat{P}^{ABCD}_a V^a + 2{\rm i}\sqrt{2}\,
\bar{\chi}^{[ABC}\psi^{D]}+\frac{3}{2}{\rm
i}\sqrt{2}\,\Omega^{[CD}\chi^{AB]E}\psi_E
\end{eqnarray}
where the graviphoton field strength $\mathcal{H}_{AB|ab}$ is
defined as
\begin{equation}
\mathcal{H}^{AB}_{ab}\equiv\mathbb{L}^{IJAB}F_{IJ\vert
ab}+\mathbb{L}_{I\alpha}^{~~AB}B^{I\alpha}_{ab}
\end{equation}
and the tensors $T_{AB}(\phi),\,A^D_{ABC}(\phi)$ are defined as
\footnote{$\left[\dots\right]\!\vert$ denotes the symplectic
traceless antisymmetrization.}
\begin{equation}
T_{AB}=T^C_{~ACB},~~A^D_{ABC}=T^D_{\left[ABC\right]\vert}\,.
\end{equation}
From the definitions (\ref{defcurvatures}) and the
parametrizations (\ref{bianchi}), applying the general procedure
described above (\ref{lie}) one immediately derives the
supersymmetry transformation laws of the physical fields (modulo
bilinears in the dilatinos):
\begin{eqnarray}
\delta V^a_{\mu} &=& - {\rm i} \bar{\varepsilon}^A \gamma^a\psi_{\mu A}  \\
\delta \psi_{A \mu}  &=&{\mathcal{D}}_\mu \varepsilon_A - g
{2\over 45}T_{AB}\gamma_\mu \varepsilon^{B} +  {2\over 3}
\mathcal{H}_{AB|\nu\mu} \gamma^{\nu} \varepsilon^B -{1\over
12}\mathcal{H}_{AB}^{\nu\rho} \gamma^{\lambda\sigma}
\varepsilon^B \epsilon _{\mu\nu\rho\lambda\sigma} \nonumber\\\nonumber\\
&& +{3{\rm i}\over 2\sqrt{2}}
\chi_{ABC}\bar{\varepsilon}^{B}\psi_{\mu}^C - {{\rm i}\over
2\sqrt{2}} \gamma_\nu\chi_{ABC} \bar\varepsilon^B\gamma^\nu
\psi_\mu^C + \mathcal{O} (\chi^2)  \\\nonumber\\
\delta \chi_{ABC} &=& {1\over\sqrt{2}} gA^D_{ABC}\varepsilon_D
+\sqrt{2} \hat{P}_{ABCD|i}\partial_\nu\phi^i
\gamma^\nu\varepsilon^D - {3\over 2\sqrt{2}}
\mathcal{H}_{[AB|\mu\nu}\gamma^{\mu\nu}\varepsilon_{C]} \nonumber\\\nonumber\\
&& - \frac{1}{2\sqrt{2}}
\Omega_{[AB}\mathcal{H}_{C]D|\mu\nu}\gamma^{\mu\nu}\varepsilon^D +
\mathcal{O} (\chi^2) \\\nonumber\\
\delta A_{IJ|\mu} &=& \mathbb{L}^{-1}_{ABIJ}\left[ {{\rm
i}\over\sqrt{2}} \bar{\chi}^{ABC}\gamma_\mu\varepsilon_C
+2{\rm i}\bar\varepsilon^A\psi^B_\mu \right]\\\nonumber\\
\delta B^{I\alpha}_{\mu\nu} &=& \mathbb{L}^{I\alpha}_{~~AB}\left[
- 2{\rm i} g \bar{\varepsilon}^A\gamma_{[\mu}\psi^B_{\nu]}
- {{\rm i}\over 2\sqrt{2}}g\bar{\chi}^{ABC}\gamma_{\mu\nu}\varepsilon_C\right]\nonumber\\\nonumber\\
&&+2\mathcal{D}_{[\mu}\left[ \mathbb{L}^{-1~I\alpha}_{AB}\left(
2{\rm i}\bar\varepsilon^A \psi^B_{\nu]} +{{\rm i}\over\sqrt{2}}
\bar{\chi}^{ABC}\gamma_{\nu]}\varepsilon_C\right)\right]
\label{beq}
 \\\nonumber\\
\hat{P}^{ABCD}_{,i}\delta \phi^i &=& 2{\rm i}\sqrt{2}\,
\bar{\chi}^{[ABC}\varepsilon^{D]}+\frac{3{\rm
i}}{\sqrt{2}}\,\Omega^{[CD}\chi^{AB]E}\varepsilon_E
\end{eqnarray}
which do indeed coincide with the corresponding formulas in
\cite{gunwar}.
\par
The B--I also give the equations of motion, which are the same as
in \cite{gunwar}. At this point, we have in principle all the
dynamical information about the theory without constructing the
Lagrangian. However, as mentioned in the previous section, in
order to get the equations of motion and the scalar potential, the
easiest way is to derive them from the superspace Lagrangian,
which, in the case of semisimple gaugings exist and we have
determined to be (up to four--fermions terms):
\begin{eqnarray}
\mathcal{L} &=& \frac{1}{24} R^{ab}\wedge V^c \cdots\wedge V^e
\epsilon_{abcde} -{{\rm i}\over 4} \bar\psi_A\gamma_{ab}
\wedge\rho^A\wedge V^a \wedge V^b +
 \nonumber\\
 &+&\frac{\rm i}{288}\bar\chi^{ABC}\gamma^a D
\chi_{ABC}\wedge V^b\cdots\wedge V^e \epsilon_{abcde} +
\nonumber\\
&-&{1\over 24} \mathcal{H}^{ab}_{AB}\left[\mathcal{H}^{AB} -
\frac{{\rm i}}{\sqrt{2}}\bar\chi^{ABC}\gamma_\ell\psi_C \wedge
V^\ell \right]\wedge V^c\cdots \wedge V^e \epsilon_{abcde}+
\nonumber\\
 &+& \frac{1}{288}
P^{ABCD\vert a}\left[ \ P_{ABCD} - 2\sqrt{2}{\rm i}
\bar\chi_{ABC}\psi_{D}\right] \wedge V^b\cdots \wedge V^e
\epsilon_{abcde}+
 \nonumber\\
&+&\left[{1\over 960} \mathcal{H}_{\ell m }^{AB}\mathcal{H}^{\ell
m}_{AB} -\frac{1}{2880} P^{ABCD}_{\ell}P_{ABCD}^\ell
\right]V^a\wedge \cdots\wedge V^e \epsilon_{abcde}+
\nonumber\\
&+&\mathcal{H}_{AB}\wedge\Bigl[\frac{\rm i}{2}
 \bar\psi^A\wedge\gamma_a\psi^B V^a + \frac{\rm i}{4\sqrt{2}}
\bar\psi_C\wedge\gamma_{ab} \chi^{ABC} V^a \wedge V^b +
\nonumber\\
&+& \frac{\rm i}{48}\bar\chi^{ALM}\gamma^{ab}\chi^B_{~~LM} V^c
\cdots \wedge V^e \epsilon_{abcde}
\Bigr]\nonumber\\
&+& \frac{\rm i\sqrt{2}}{72}
P_{ABCD}\bar\chi^{ABC}\gamma^{ab}\psi^D\wedge V^c \cdots
\wedge V^e \epsilon_{abcde} +\nonumber\\
 &-&
\frac{{\rm i}}{180}g \left[ T_{AB}\bar\psi^A \gamma^{ab}\wedge
\psi^B
 +\frac{5}{8}A^A_{~BCD}\bar\psi_{A}
 \gamma^a\chi^{BCD}\wedge V^b\right]\wedge V^c \cdots\wedge V^e \epsilon_{abcde}+
 \nonumber\\
&+& g^2\frac{1}{120}\left[{6\over (45)^2}T_{AB}T^{AB}-{1\over
  96}A_{ABCD}A^{ABCD}\right]V^a\wedge \cdots\wedge V^e
  \epsilon_{abcde}+
\nonumber\\
&-& \frac{1}{24}\epsilon^{IJKLMN} \Bigl[ {\cal F}_{IJ}\wedge {\cal
F}_{KL} \wedge A_{MN}
 +  g\eta^{PQ}{\cal F}_{IJ}\wedge A_{KL} \wedge A_{MP}\wedge A_{QN}
 +\nonumber\\
 &+& \frac{2}{5}
  g^2\eta^{PQ} \eta^{RS}A_{IJ}\wedge A_{KP} \wedge A_{QL}\wedge A_{MR}\wedge A_{SN}
 \Bigr]+
 \frac{1}{2g}\eta_{IJ}\epsilon_{\alpha\beta} {\cal B}^{I\alpha}\wedge
 H^{J\beta}
  + \mathcal{L}_{4-f}\nonumber\\
  \label{lagrangian}
\end{eqnarray}
where the two--forms ${\cal H}_{AB}$ are defined by
\begin{equation}
\mathcal{H}^{AB}\equiv\mathbb{L}^{IJAB}{F}_{IJ}+
\mathbb{L}_{I\alpha}^{~~AB}{B}^{I\alpha}
\end{equation}
and we have introduced, in the Chern--Simons contributions,
\begin{eqnarray}
{\cal F}_{IJ} &\equiv & F_{IJ} + {\rm
i}\mathbb{L}^{-1}_{ABIJ}\bar\psi^A \wedge  \psi^B \\
{\cal B}^{I\alpha} & \equiv & {B}^{I\alpha} + {\rm
i}\mathbb{L}^{-1~~I\alpha}_{AB}\bar\psi^A \wedge\psi^B\,.
\end{eqnarray}
\par
The projection of (\ref{lagrangian}) onto space--time coincides
with eq. (4.15) in \cite{gunwar}.
\subsection{The problem of the two--forms}
It is a known fact \cite{PPV}, \cite{gunwar} that in order to
consistently gauge the $\mathcal{N}=8$ theory, one has to dualize
the vectors transforming in the $(\bf{6},\bf{2})$ of
$\mathrm{SO(p,q)}\times \mathrm{SL}(2)$ to massive two-forms
obeying the self-duality constraint:
\begin{equation}
B^{I\alpha |\mu\nu} = m
\epsilon^{\mu\nu\rho\sigma\lambda}{\cal{D}}_\rho
B^{I\alpha}_{\rho\sigma\lambda}\,. \label{self}
\end{equation}
In the geometric formulation of the theory, this need for
dualization emerges in a completely natural way. Indeed, let us
start by considering the 12 vectors $A^{I\alpha}$. There is no way
known to reconcile their abelian gauge invariance with their
non-trivial transformation under the gauge group $\mathcal{G}$.
Indeed, given the superspace curvatures
\begin{equation}
 DA^{I\alpha} \equiv dA^{I\alpha}+g(G^{KL})^I_{\,J}A_{KL}\wedge A^{J\alpha}
\end{equation}
it follows that the corresponding Bianchi identity contains a
term:
\begin{equation}
DDA^{I\alpha} = g(G^{KL})^I_{\,J}{\cal F}_{KL}\wedge
A^{J\alpha}+\dots
\end{equation}
where the vectors $A^{J\alpha}$ appear naked. Under such
conditions we cannot write a rheonomic parametrization of the
curvatures solving the Bianchi identities and containing as only
possible terms monomials in vielbein and gravitino with
coefficients expressed in terms of gauge invariant space--time
curvature components. Hence we have a clash between supersymmetry
and the $12$ abelian gauge invariances needed to keep the vectors
$A^{J\alpha}$ massless. On the other hand, making them massive
would destroy the equality of the Bose and Fermi degrees of
freedom. Hence, in the gauged case where the $12$ vectors
$A^{J\alpha}$ acquire a non--trivial transformation under the
non--abelian gauge symmetry there is no way of fitting these
fields into a  consistent supersymmetric theory. The way out, as
it was discussed in \cite{gunwar}, is to interpret them as the
duals of massive two-forms $B^{I\alpha}$ \footnote{In five
dimensions the Hodge dual of a two--form field strength is a
three--form field strength:
$\epsilon^{\mu\nu\rho\lambda\sigma}D_\lambda A_{\sigma}= D^\mu
B^{\nu\rho}$.}, obeying a self--duality constraint which halves
their degrees of freedom. This construction emerges naturally in
the rheonomic formulation based on Free Differential Algebras. In
this context, one has to introduce superspace curvatures for the
two--forms (see eq.s (\ref{defcurvatures})) generalizing the
Maurer--Cartan equations to a Free Differential Algebra
\cite{castdauriafre}, \cite{Fre:1984pc}. At first sight it seems
that we cannot escape from the problem described above, that
affects the vectors $A^{I\alpha}$:  indeed Bianchi identities do
contain the naked fields $B^{I\alpha}$. Yet we can successfully
handle this fact by considering the $B^{I\alpha}$ not as gauge
potentials (that is, 2-forms defined modulo $1$--form gauge
transformations), but as physical fields, with their own explicit
parametrization (see equation (\ref{rheoB})) \footnote{The same
happens to matter two--form fields coupled with $\mathcal{N}=2$
supergravity \cite{Ceresole:2000jd}.}. In this way, the two--forms
loose their gauge freedom and become massive, as it can be found
by solving the Bianchi identities. In fact, the Bianchi identities
imply also  the field equations of the two--forms. In the
 $\bar{\psi}^A\wedge\gamma^a\psi_A$ sector of the $H^{I\alpha}$ sector we get the
 following constraint (modulo bilinears in the dilatinos):
\begin{equation}
\label{eqmotB} D_{[a}B^{I\alpha}_{bc]}= -{1\over 12} g
\mathbb{L}^{I\alpha}_{~~AB}\mathcal{H}^{AB~|de}\varepsilon_{abcde}
-{\rm
i}\sqrt{2}\mathbb{L}^{-1~I\alpha}_{AB}\bar{\chi}^{ABC}\gamma_{[a}\rho
_{bc]C}\,.
\end{equation}
This is the self-duality constraint on the two-forms, that halves
the number of  their degrees of freedom, while gives them a mass
$g$. Note that the algebra underlying this theory is a free
differential algebra \cite{sull,castdauriafre,Fre:1984pc}.
However, since the $B^{I\alpha}$ transform in a non trivial
representation of the gauge group, it can be shown (as it follows
from a theorem by Chevalley--Eilenberg \cite{sull}) that the left
hand side of their rheonomic parametrization at
$B^{I\alpha}_{ab}=0$ is  a trivial cohomology class of the algebra
spanned by the other one--forms. Said in simpler terms, in the
vacuum (where the superspace curvatures are zero) the two--forms
are not independent fields, rather they are names given to certain
bilinears in the gravitino fields:
\begin{equation}
B^{I\alpha}=-{\rm
i}\delta^{I\alpha}_{AB}\bar{\psi}^A\wedge\psi^B\,.
\end{equation}
There is a drastic algebraic difference between these $2$--forms
and the $p$--forms that appear in most higher dimensional
supergravities. In the mathematical language of Sullivan
\cite{sull} this  has to do with the distinction between {\it non
trivial minimal free differential algebras} and {\it trivial
contractible algebras}. In the first case the exterior derivative
of a $p$--form is equated to a non--trivial cohomology class of
the superalgebra spanned by the other forms, namely to a
polynomial in the remaining  $1$--forms that cannot be written as
the derivative of any other such polynomial. In the second case
the derivative of the $p$--form is equated to a trivial class.
True $p$--form gauge fields occur only when  the vacuum free
differential algebra (that at zero curvature) is minimal. On the
other hand, if the vacuum free differential algebra is
contractible then there are no true $p$--form gauge fields since
they can be traded for an expression in terms of the other
$1$--forms. As it was shown in \cite{Fre:1984pc} the contractible
generators of a free differential algebra are anyhow associated
with the concept of curvatures. Indeed when a minimal algebra is
deformed by the introduction of curvatures it becomes
contractible. So the self--duality between the field--strengths
(=curvatures) and the 2--form potentials acquires in this language
a natural explanation. It is just the signal that the FDA is
contractible. This, in line with Chevalley-Eilenberg theorem is
due to the semisimple character of the super--Lie algebra of which
the FDA is the extension.
\par
We  stress that in  supergravity theory one usually deals  with
massless $p$--forms. This reduces their degrees of freedom by
means of gauge invariance. When we gauge the theory, it often
occurs that the $p$--forms become massive and some other mechanism
has to intervene in order to reduce their number of degrees of
freedom and keep the balance between fermions and bosons. This
mechanism can be either self--duality, as it happens in our case
and in seven dimensional supergravity, or the so--called
anti--Higgs mechanism, as it happens in gauged $F(4)$ gauged
supergravity in six dimensions \cite{F4} \footnote{The former can
occur only in odd dimensional space--times \cite{townsend}.}. Let
us note that the self--duality mechanism is a relation  between a
form and its field strength, stating in this way the triviality of
the cohomology related to that form. As it follows from the
theorem  in \cite{sull}, this necessarily happens when the forms
are in some non trivial representation of the gauge group. On the
contrary, we expect the anti--Higgs mechanism, which implies a non
trivial cohomology for the form, to be present only when the form
is a gauge singlet.
\section{Gauging the non--semisimple $\mathrm{CSO}\mathrm{(p,q,r)}$ groups}
In the original papers \cite{gunwar,PPV} gauged versions of
five--dimensional maximal supergravity were constructed where the
gauge group is either $\mathrm{SO(6)}$ or one of its non compact
forms $\mathrm{SO(p,q)}$ (with $p+q=6$). This is similar to what
happens in four--dimensions where the semisimple gaugings of
$\mathcal{N}=8 $ supergravity are based on all the groups
$\mathrm{SO(p,q)}$ with $p+q=8$. In fact in that theory there is
an additional series of interesting non--semisimple gaugings based
on contracted algebras $CSO(p,q,r)$ (with $p+q+r=8$) whose notion
was introduced by Hull \cite{hull,hull2} and whose classification
was shown to be exaustive in \cite{noi4D}. It is quite natural to
expect that such non--semisimple gaugings exist also in
five--dimensions with $p+q+r=6$. However they were not constructed
in \cite{gunwar,PPV} because of the subtle features related with
the problem of two--form gauge fields. This problem being
naturally solved in the Free Differential Algebra rheonomic
approach we are tempted to argue that the
$\mathrm{CSO}\mathrm{(p,q,r)}$ can be constructed in this
framework. This is indeed true as we explicitly show below. Indeed
the catch point is that the number of vectors dualized to
two--forms is not fixed to $12$ as in the semisimple gaugings
rather it is variable. In the non semisimple
$\mathrm{CSO}\mathrm{(p,q,r)}$ case we have $12-r$ two--forms and
$15+r$ 1--forms. However $r$ of these latter do not gauge any
transformation with non trivial action on the other fields, in
other words they are associated with {\it central charges}. The
price to be paid for that seems to be that a lagrangian
formulation is not available for these theories, all the dynamical
information being however encoded in the solution of the Bianchi
Identities.
\subsection{The $\mathrm{CSO}\mathrm{(p,q,r)}$ algebras}
We begin with a short description of the contracted algebras and
in the next subsection we explain how they are gauged.
\par
The generators of $\mathrm{SO(p,q)}$ (with $p+q=n$) in the vector
representation are
\begin{equation}
(G^{IJ})^K_{~~L}=\delta^{[K}_J\eta^{L]I}~~~~I,J,K,L=1,\dots,n\,,
\end{equation}
where
\begin{equation}
\eta^{IJ}\equiv\,{\rm
diag}(\overbrace{1,\dots,1}^p,\overbrace{-1,\dots,-1}^q)\,.
\end{equation}
They satisfy
\begin{equation}
[G^{IJ},G^{KL}]=f^{IJ,KL}_{MN}G^{MN}
\end{equation}
where
\begin{equation}
\label{structconst}
f^{IJ,KL}_{MN}=-2\delta^{[I}_{[M}\eta^{J][K}\delta^{L]}_{N]}\,.
\end{equation}
Their generalization, studied by Hull in the context of
supergravity \cite{hull},\cite{hull2} are the algebras
$\mathrm{CSO}(p,q,\!r)$ with $p+q+r=n$, defined by the structure
constants (\ref{structconst}) with
\begin{equation}
\eta^{IJ}\equiv\,{\rm
diag}(\overbrace{1,\dots,1}^p,\overbrace{-1,\dots,-1}^q,
\overbrace{0,\dots,0}^r)\,. \label{metriccpqr}
\end{equation}
Decomposing the indices
\begin{equation}
I=(\bar{I},\hat{I})~~~~\bar{I}=1,\dots,p+q,~\hat{I}=p+q+1,\dots,n\,,
\end{equation}
we have that $G^{\bar{I}\bar{J}}$ are the generators of
$\mathrm{SO(p,q)}\subset \mathrm{CSO}\mathrm{(p,q,r)}$ , while the
$r(r-1)/2$ generators $G^{\hat{I}\hat{J}}$ are central charges
\begin{equation}
[G^{\bar{I}\hat{J}},G^{\bar{K}\hat{L}}]={1\over
2}\eta^{\bar{I}\bar{K}} G^{\hat{J}\hat{L}}\,. \label{nonnull}
\end{equation}
They form an abelian subalgebra, and
\begin{equation}
\mathrm{SO(p,q)}\times \mathrm{U(1)}^{r\left(r-1\right)\over
2}\subset \mathrm{CSO(p,q,r)}\,.
\end{equation}
Notice that $\mathrm{\mathrm{CSO}(p,q,1)}=\mathrm{ISO(p,q)}$. In
the vector representation, the generators of the central charges
are identically null
\begin{equation}
(G^{\hat{I}\hat{J}})^K_{~L}=0\,, \label{zerogen}
\end{equation}
while
\begin{equation}
(G^{\hat{I}\bar{J}})^K_{~L}={1\over
2}\delta^{\hat{I}}_L\eta^{\bar{J}K}\neq 0\,.
\end{equation}
It is worth noting that the Killing metric of $\mathrm{SO(p,q,r)}$
is
\begin{equation}
K^{IJ,KL}=f^{IJ,MN}_{PQ}f^{KL,PQ}_{MN}=-6\eta^{K[I}\eta^{J]L}\,.
\end{equation}
This notation is redundant, because the adjoint representation is
$n(n-1)/2$ dimensional. In the proper basis,
\begin{equation}
K^{IJ,KL}_{_{I<J,\,K<L}}=-3\eta^{IK}\eta^{JL}\,.
\end{equation}
This is a diagonal matrix of dimension $n(n-1)/2$, with components
$\eta^{II}\eta^{JJ}$. In general, the real sections of a given
group (in this case, $D_3$) are characterized by the signature of
the Killing metric\footnote{for non semisimple groups, by
signature we mean the number of positive, negative and null
components of the matrix $\eta$ in its diagonal form}. We see
that, for the $\mathrm{CSO}\mathrm{(p,q,r)}$ algebras, the
signature of the Killing metric is equivalent to the signature of
the matrix $\eta^{IJ}$. This explains why this tensor can give an
intrinsic characterization of such groups. Notice that a similar
result was found, with a different procedure, while studying the
gaugings of $\mathcal{N}=8$ supergravity in four space--time
dimensions \cite{noi4D}.
\subsection{The contracted gaugings}
As announced above the gauged versions of $\mathcal{N}=8$, $D=5$
supergravity constructed in \cite{gunwar}, \cite{PPV} and based on
a semisimple choice of the gauge group
$\mathcal{G}=\mathrm{SO(p,q)}~(p\!+\!q\!=\!6)$ can be further
generalized to the non--semisimple gauge groups
$\mathcal{G}=\mathrm{CSO}\mathrm{(p,q,r)}~(p\!+\!q\!+\!r\!=\!6)$.
\par
The new gaugings can be obtained by taking for the matrix
$\eta^{IJ}$ the definition (\ref{metriccpqr}), with some null
entries on the diagonal. Let us discuss the consequences of  this
in the theory, in order to see if any pathology occurs. One has
\begin{equation}
(G^{KL})^{\hat{I}}_{~J}=\delta^{[K}_J\eta^{L]\hat{I}}=0\,,
\end{equation}
so the covariant derivative of a {\sl contravariant field}
(\ref{covGder}), along the contracted directions, reduces to the
ordinary  $\mathrm{USp(8)}$--covariant derivative:
\begin{equation}
DV^{\hat{I}}= \nabla
V^{\hat{I}}+g(G^{KL})^{\hat{I}}_{~J}A_{KL}\wedge V^J =\nabla
V^{\hat{I}}\,.
\end{equation}
This, however, does not happen for the covariant derivative of a
{\sl covariant field}:
\begin{equation}
DV_{\hat{I}} \equiv \nabla
V_{\hat{I}}-g(G^{KL})^J_{~\hat{I}}A_{KL}\wedge V^J= \nabla
V_{\hat{I}}-g\eta^{\bar{L}\bar{J}}A_{\hat{I}\bar{L}}\wedge
V_{\bar{J}}\,.
\end{equation}
The abelian vectors $A_{\hat{I}\hat{J}}$ do not appear in the
covariant derivatives. Because of (\ref{nonnull}), in the field
strengths
\begin{equation}
\mathcal{F}_{IJ}=dA_{IJ}+{1\over 2}f^{KL,MN}_{IJ}A_{KL}\wedge
A_{MN}
\end{equation}
the last term is present even for the vectors of the abelian
subgroup. Let us consider now the most subtle part of the theory:
the two--forms. Along the contracted directions, one has
\begin{equation}
\mathbb{L}^{\hat{I}\alpha AB}\equiv
\eta^{\hat{I}J}\varepsilon^{\alpha\beta}\mathbb{L}_{J\beta}^{~~AB}=0\,,
\end{equation}
so that the rheonomic parametrization (\ref{rheoH}) of
$H^{\hat{I}\alpha}$ becomes
\begin{equation}
H^{\hat{I}\alpha}=H^{\hat{I}\alpha}_{abc}V^a\wedge V^b\wedge
V^c\,. \label{rheoHcontr}
\end{equation}
The corresponding Bianchi identity reads:
\begin{equation}
DH^{\hat{I}\alpha}=0\,,
\end{equation}
and, substituting  back the parametrization (\ref{rheoHcontr}),
one finds
 $H^{\hat{I}\alpha}_{abc}=0$. Hence we have:
\begin{eqnarray}
0&=&H^{I\alpha}=d\left[B^{\hat{I}\alpha}+{\rm i}
\mathbb{L}^{-1~\hat{I}\alpha}_{AB}\bar{\psi}^A\wedge\psi^B\right]
+ g(G^{KL})^{\hat{I}}_{~J}A_{KL}\wedge
\left[B^{\hat{I}\alpha}+{\rm i}
\mathbb{L}^{-1~\hat{I}\alpha}_{AB}\bar{\psi}^A\wedge\psi^B\right] =\nonumber\\\nonumber\\
&=&d\left[B^{\hat{I}\alpha}+{\rm i}
\mathbb{L}^{-1~\hat{I}\alpha}_{AB}\bar{\psi}^A\wedge\psi^B\right]
\,. \label{bianchiAhatI}
\end{eqnarray}
The solution of this equation is
\begin{equation}
B^{\hat{I}\alpha}=\mathcal{B}^{\hat{I}\alpha}-{\rm i}
\mathbb{L}^{-1~\hat{I}\alpha}_{AB}\bar{\psi}^A\wedge\psi^B
\end{equation}
with
\begin{equation}
\mathcal{B}^{\hat{I}\alpha}\equiv dA^{\hat{I}\alpha}\,.
\label{abelianvectordef}
\end{equation}
In other words, the Bianchi identities of the two--forms
corresponding to the contracted direction (the
$B^{\hat{I}\alpha}$) are cohomologically trivial, so that these
fields are actually field strengths of one--form fields
(\ref{abelianvectordef}), having a $\mathrm{U(1)}$ gauge
invariance, as argued in \cite{gunwar}. Let us stress that this
explicit calculation performed in the rheonomy formalism shows
that there are no consistency conflicts between the two types of
gauge invariances, and therefore no need arises to introduce
massive vectors as proposed in \cite{gunwar}. Indeed,  in the FDA
rheonomic approach we  see in a transparent way where the
consistency conflicts arise and how they are solved. Summarizing
it goes as follows. When a vector field is charged with respect to
the gauge group, but does not gauge any generator of the gauge
algebra it appears naked in its own Bianchi Identity. This
requires dualization to a two--form and the replacement of gauge
invariance with self duality as a mean to reduce the number of
degrees of freedom. On the other hand  when a contraction is
performed on some direction $\hat{I}$, in the Bianchi identity of
the fields $A^{\hat{I}\alpha}$ (\ref{bianchiAhatI}) the naked
gauge fields disappear. Therefore, the two gauge invariances are
not inconsistent, and the corresponding vectors can stay massless.
Note that in this case the Bianchi identities look very different
from those along the non-contracted directions. Now the
self--duality constraint disappears and the halving of degrees of
freedom is due to the recovered $\mathrm{U(1)}$ gauge symmetry.
\par
In this way we have found new gauged $D=5~\mathcal{N}=8$
supergravities, with $(12-r)$ two--forms, $(15+r)$ one--forms, and
gauge group $\mathrm{CSO}\mathrm{(p,q,r)}$. It is worth noting
that the $r$ vectors $A^{\hat{I}\alpha}$ are coupled with the
other fields, even if they don't gauge anything, and so are the
abelian  vectors $A_{\hat{I}\hat{J}}$. Indeed,
\begin{equation}
\mathcal{H}^{AB}_{ab}=\mathbb{L}^{IJAB}F_{ab\vert IJ}+
\mathbb{L}_{\bar{I}\alpha}^{~~AB}B^{\bar{I}\alpha}_{ab}+
\mathbb{L}_{\hat{I}\alpha}^{~~AB}B^{\hat{I}\alpha}_{ab}
\end{equation}
and $\mathcal{H}^{AB}_{ab}$ does appear in the equation of motion
of the two--form (\ref{eqmotB}) along the non--contracted
direction, which we have derived from the B-I and doesn't change
in the contracted gaugings.
\par
The susy transformatons rules for the new theories are obtained by
substituting, for the contracted directions $\hat I$, eq.
(\ref{beq}) with:
\begin{equation}
\delta A^{\hat I\alpha|\mu} = \mathbb{L}^{-1~\hat
I\alpha}_{AB}\left[ {{\rm i}\over\sqrt{2}}
\bar{\chi}^{ABC}\gamma_\mu\varepsilon_C +2{\rm
i}\bar\varepsilon^A\psi^B_\mu \right]
\end{equation}
 all other transformation laws remaining unchanged.
 \par
 The new theories are completely sensible and well
 defined, however it seems that there is not a lagrangian formulation
 of them.
 Indeed, by looking at (\ref{lagrangian}) describing the semisimple
 gaugings, one can see that it involves the inverse matrix
 $\eta_{IJ}$ \footnote{We remind that the definition of the
$CSO(p,q,r)$ generators in the vector representation involves the
matrix $\eta^{IJ}$.} in the Chern-Simons term for the two-forms.
This matrix is not well defined for the contracted directions
$\hat I$. The corresponding terms cannot therefore be present in
the lagrangian. As a consequence, it seems to us that there is no
way to write down new terms in the lagrangian substituting the
badly defined ones in a covariant and gauge-invariant way.
Technically, the obstruction is related to the fact that that
Chern-Simons contribution was necessary for the vanishing of the
variation of the lagrangian with respect to the vectors $A_{IJ}$,
by use of the cubic invariant of $E_{6(6)}$. One could try to add
by hand in the lagrangian the missing terms which complete the
cubic invariant (even if their meaning would be quite obscure) but
they would lead to not-gauge invariant field equations for the new
vectors $A^{\hat I \alpha}$. Otherwise, one could try to interpret
the badly defined terms in the lagragian as the dominant ones
(since they are now infinite) through some appropriate scaling
limit. However, since the connection between the covariant and
contravariant representions for the $\hat I$ directions is lost,
we were not able to implement in a covariant way the field
equations through appropriate terms to be added to the lagrangian.
We postpone to the concluding section some comments about this
fact.
 \par
\subsection{Conclusions}
\par
In this paper, we have proven the existence of $D=5~{\cal N}=8$
supergravities where the non--semisimple $CSO(p,q,r)$ algebras are
gauged and this is a novelty since it was so far unclear in the
literature whether this could be done or not. Of these theories,
at the present stage, we possess the supersymmetry algebra whose
closure implies the field equations. According to what we
explained in section $4.2$, the lagrangian formulation of these
theories probably does not exist. A possible physical argument to
motivate this situation that was technically illustrated above is
the following. The existence of $r$ extra neutral vectors besides
the $15$ charged ones implies a sort of Hodge dualization for the
corresponding two--forms. Specifically, what happens here is that
the field strength $H^{[3]}$ for $r$ of the $B^{[2]}$ fields is
identically zero, so that we have to interpret the $B^{[2]}$
themselves as field strengths of new gauge vectors $A^{[1]}$. In
other words, we have traded $r$ ''electric'' two--form fields
$B^{[2]}$ for just as many ''magnetic''  one--forms $A^{[1]}$. In
view of this, it is not too surprising if the $15+r$ vectors are
not mutually local, which would be necessary to admit a common
lagrangian description.
\par
However, all the information on the dynamics of these theories are
encapsulated in the rheonomic solution of the B-I that we have
presented. As explained in section $3.1$, to extract the explicit
form of the field equations and of the potential it suffices to
analyze the remaining sectors of the B-I, through somewhat lengthy
calculations, whose result, however, is a priori guaranteed.
\par
Although these calculations and a full--fledged analysis of the
field equations and their solutions are postponed to a forthcoming
publication \cite{next}, we can briefly anticipate some
considerations on the scalar potential that, most presumably, has
the same form, in terms of the fermionic shift and gravitino mass
matrix, as in the compact case \cite{gunwar}:
\begin{equation}
P=-g^2\left[{2\over 675}T_{AB}T^{AB}-{1\over
96}A_{ABCD}A^{ABCD}\right]\,. \label{potential}
\end{equation}
Indeed, expression (\ref{potential}) follows from the general
structure of supersymmetry Ward identities one finds in gauged
supergravities.
\par
As it has been shown in \cite{gunwar}, the number of
supersymmetries preserved by a constant scalar configuration
$\phi_0$ is given by the number of eigenvalues $\mu$ of
\begin{equation}
W_{AB}(\phi_0)\equiv {4\over 15}T_{AB}(\phi_0)
\end{equation}
such that
\begin{equation}
\left|\mu\right|=\sqrt{-{3\over g^2}P(\phi_0)}\,.
\end{equation}
The scalars are in the $\mathcal{G}$ representations
\begin{equation}
\bf 20\oplus 10\oplus\bar{10}\oplus 1\oplus\bar{1}\,.
\end{equation}
The potential is invariant under all the local and global
symmetries of the theory, which in this case are
$\mathcal{G}\times \mathrm{SL}(2,\mathbb{R})$. To look for minima
of the potential (\ref{potential}) one can use Warner's
observation \cite{schur} that, given a subgroup
$\mathcal{G}'\subset\mathcal{G}\times \mathrm{SL}(2,\mathbb{R})$,
and given the submanifold of the scalar manifold
$\Sigma'\subset\Sigma$ invariant under the action of
$\mathcal{G}'$, then the minima of $\Sigma'$ are also minima of
$\Sigma$. We can therefore restrict the potential to subsets of
the scalars, and search their minima. In particular, the scalars
invariant under $\mathcal{G}'=\mathrm{U(1)} \subset
\mathrm{SL}(2,\mathbb{R})$ are those in the $\bf 20$, that
parametrize the coset $\mathrm{SL}(6,\mathbb{R})/\mathrm{SO(6)}
\subset \mathrm{E_{6(6)}} /\mathrm{USp(8)}$. One can study the
minima restricted to these scalars. In this case we can use a
simple formula to find the potential, found in \cite{gunwar} for
the $\mathrm{SO(p,q)}$ case
\begin{equation}
\label{potential20} P=-{g^2\over
32}\left[\left(\eta^{IJ}M_{IJ}\right)^2-2\left(\eta^{IJ}M_{IJ}
\eta^{KL}M_{KL}\right)\right]\,,
\end{equation}
where
\begin{equation}
M_{IJ}=S_I^{~K}S_J^{~K}
\end{equation}
and $S_I^{~K}$ are the $\mathrm{SL}(6,\mathbb{R})$ generator
associated with the scalars we are considering. Furthermore,
\begin{equation}
W_{AB}=-{1\over 4}\Omega_{AB}\eta^{IJ}M_{IJ}\,. \label{W20}
\end{equation}
The formulas (\ref{potential20}), (\ref{W20}) remain valid for the
contracted theories, where the $\eta^{IJ}$ have zero eigenvalues.
However this procedure is not very useful to find non--maximally
supersymmetric minima of the potential. Indeed, since we have
$W_{AB}\propto\Omega_{AB}$ all $W_{AB}$ eigenvalues $\mu$  are
equal in modulus and differ only for the phase. This implies that
restricting our attention to the scalars in the $\bf 20$
representation of $\mathrm{SL(6,\mathbb{R})}$ either there is
$\mathcal{N}=8$ supersymmetry, or there is no supersymmetry. Yet a
minimum with $\mathcal{N}=8$ supersymmetry should be invariant
under  $SU(4)=\mathrm{SO(6)}$ and this occurs only when all the
scalars are set to zero and only for the $\mathrm{SO(6)}$ theory.
Hence, in the quest for other supersymmetric minima,  one must
necessarily consider the scalars in the $\bf
10\oplus\bar{10}\oplus 1\oplus\bar{1}$, as it was done for the
only other known supersymmetric minimum, found \cite{N2minimum} in
the $\mathrm{SO(6)}$ theory.
\par
Such an analysis is quite involved and it is beyond the scope of
the present paper. However it is a very interesting and
challenging problem that we postpone to the already mentioned
future investigations. We just note that the contracted gaugings
yield a non supersymmetric vacuum with zero cosmological constant
\footnote{The same phenomenon occurs in the four dimensional case
with the $\mathrm{CSO}(2,0,6)$ gauging \cite{hull2}.}. In the
$\mathrm{CSO}(2,0,4)$ gauging, the potential, restricted to the
scalars invariant under $\mathrm{SO(2)}\times \mathrm{U(1)}^6$,
vanishes identically. Because of Warner's argument \cite{schur}
this implies  that these scalars correspond to a minimum of the
whole potential, which has zero cosmological constant.
\par
Summarizing, in this paper we have shown that there exist new
non--semisimple gauged supergravities in five dimensions that are
potentially very interesting in the quest for brane--worlds and
the DW/QFT correspondence. In perspective, it is very interesting
to find an interpretation of these theories as the supergravities
describing the near--brane geometry for suitable stringy branes.
\section*{Acknowledgements}
One of the authors (L.A.) warmly thanks the Physics Department of
the K.U.Leuven where most of this work has been carried on. L.G.
is supported in part by the ``Actions de Recherche Concert{\'e}es"
of the ``Direction de la Recherche Scientifique - Communaut{\'e}
Fran{\c c}aise de Belgique", by IISN - Belgium (convention
4.4505.86). This work has been supported by the European
Commission TMR programme ERBFMRX-CT96-0015 and by the European
Commission RTN programme RTN1-1999-00116 in which L.A. and P.F.
are associated to Torino and L.G. is associated to K.\ U.\ Leuven.

\end{document}